\documentclass[12pt,a4paper,oneside]{article}
\usepackage[top=3cm, bottom=3cm, left=2cm, right=2cm]{geometry}
\usepackage[english]{babel}
\usepackage[utf8]{inputenc}
\usepackage[T1]{fontenc}

\usepackage{amsthm,amsmath,amssymb,mathrsfs,dsfont,mathtools}
\usepackage{bbm}
\usepackage{bm}
\usepackage{graphicx}
\usepackage[colorlinks=true, allcolors=blue]{hyperref}
\usepackage{comment}
\usepackage[all]{nowidow}
\usepackage{booktabs}
\usepackage{cleveref}
\usepackage{multirow}

\usepackage[authoryear, round]{natbib}

\newcommand{\ind}{\stackrel{\mbox{\scriptsize ind}}{\sim}}

\newcommand{\E}{\mathrm E}
\newcommand{\V}{\mathrm Var}
\newcommand{\ddr}{\mathrm{d}}
\newcommand{\indic}{\mathds{1}}

\usepackage[plain,noend]{algorithm2e}

\makeatletter
\renewcommand{\algocf@captiontext}[2]{#1\algocf@typo. \AlCapFnt{}#2} 
\def\@algocf@capt@plain{top}
\renewcommand{\algocf@makecaption}[2]{%
  \addtolength{\hsize}{\algomargin}%
  \sbox\@tempboxa{\algocf@captiontext{#1}{#2}}%
  \ifdim\wd\@tempboxa >\hsize
    \hskip .5\algomargin%
    \parbox[t]{\hsize}{\algocf@captiontext{#1}{#2}}
  \else%
    \global\@minipagefalse%
    \hbox to\hsize{\box\@tempboxa}
  \fi%
  \addtolength{\hsize}{-\algomargin}%
}
\makeatother

\NewDocumentCommand{\evalat}{sO{\big}mm}{%
  \IfBooleanTF{#1}
   {\mleft. #3 \mright|_{#4}}
   {#3#2|_{#4}}%
}

\renewcommand{\mid}{\ensuremath{\,|\,}}

\usepackage{authblk}
\date{}
\title{Large-scale entity resolution via microclustering Ewens--Pitman random partitions}
\author[1]{Mario Beraha}
\author[2]{Stefano Favaro}

\affil[1]{Department of Economics, Management, and Statistics, University of Milano--Bicocca, 20126 Milano, Italy}
\affil[2]{Department of Economics and Statistics, University of Torino and Collegio Carlo Alberto, 10134 Torino, Italy}

\providecommand{\keywords}[1]{
  \small 
  \textbf{\textit{Keywords:}} #1
  \normalsize
}

\usepackage[sf]{titlesec}
\usepackage{sectsty}
\allsectionsfont{\centering \normalfont\scshape} 

\newtheorem{theorem}{Theorem}

\newtheorem{proposition}{Proposition}
\theoremstyle{definition}

\newtheorem{definition}{Definition}
\newtheorem{remark}{Remark}

\begin{document}

\maketitle

\begin{abstract}
We introduce the microclustering Ewens--Pitman model for random partitions, obtained by scaling the strength parameter of the Ewens--Pitman model linearly with the sample size. 
The resulting random partition is shown to have the microclustering property, namely: the size of the largest cluster grows sub-linearly with the sample size, while the number of clusters grows linearly. By leveraging the interplay between the Ewens--Pitman random partition with the Pitman--Yor process, we develop efficient variational inference schemes for posterior computation in entity resolution. Our approach achieves a speed-up of three orders of magnitude over existing Bayesian methods for entity resolution, while maintaining competitive empirical performance.
\end{abstract}



\section{Introduction}

\subsection{Motivation: large--scale entity resolution}

Entity resolution (ER) is the task of identifying which noisy, duplicate or incomplete records refer to the same real-world entity. It is a critical component of data integration, underpinning a wide range of high-impact applications across many sectors. In healthcare, the inability to reliably match patient records across hospital systems remains both common and costly \citep{Pew2018,AHIMA2024}. National statistical agencies rely on large-scale record linkage to construct deduplicated population frames for census and survey operations \citep{USCensus2023}. In the private sector, Customer-360 initiatives depend on accurate identity resolution to unify fragmented customer touchpoints and deliver personalized services \citep{AWS2023,Wardwell2023}.  

In all these settings, the sample size $n$ is often on the order of tens of thousands or more, while each individual entity is typically represented with only a small number of  records. This pronounced imbalance, namely many-records-per-entity, is a defining feature of ER, and it imposes specific requirements with respect to the statistical behavior of the underlying clustering model. In particular: i) the size of the largest cluster should grow sub-linearly with $n$; and ii) the number of clusters, as well as the number of clusters of any fixed size $r\geq1$, should grow linearly with $n$. These growth conditions are collectively referred to as the microclustering property \citep{Bet22}. Clustering models that fail to satisfying this property tend to overstate uncertainty and, in practice, produce unreliable or unusable summaries of the resolved entities.

\subsection{Background and challenges}

Distributions for random partitions play a fundamental role as prior models in Bayesian clustering, most prominently the Ewens--Pitman (EP) model induced by random sampling the Pitman--Yor process \citep{Pitman95,PitmanYor97,Deblasi13}. However, a key limitation of the EP prior is that it fails to satisfy the microclustering property, as the size of the largest cluster grows linearly with the sample size $n$, i.e., on the order of $O(n)$.

Recent works have proposed alternative priors for random partitions that satisfy the microclustering property \citep{Bet16, Mil15, DiBenedetto21, Bet22}. Although these priors are theoretically well-founded, their posterior inference relies on specialized marginal Markov chain Monte Carlo (MCMC) algorithms with computational costs scaling quadratically in the sample size $n$. This computational burden makes them impractical for large-scale ER. In contrast, the EP prior benefits from simple conditional algorithms based on stick-breaking representations of the Pitman--Yor process \citep{ishwaran2001gibbs}, and it is also well-suited for efficient variational inference (VI) techniques \citep{Blei06}.

\subsection{Preview of our contributions}

We show that a scaling of the EP prior with respect to the sample size $n$ yields the microclustering property. In turn, this allows us to develop an efficient VI algorithm to perform large scale ER. The intuition behind our results originates from the work of \citet{Contardi24}, who study asymptotic properties the EP prior with strength parameter $\theta > 0$ and discount parameter $\alpha \in [0, 1)$. In particular, they show that number of clusters grows linearly with $n$ when the prior is scaled by setting $\theta=\lambda n$, for $\lambda>0$. Building on this insight, we prove that under the same scaling of the EP prior, the size of the largest cluster grows sub-linearly with $n$, while the number of clusters of any fixed size $r\geq1$ grows linearly with $n$, thus fulfilling the microclustering property. We refer to the scaled EP prior as the microclustering EP (M-EP) prior.

The M-EP prior enables the use of efficient posterior inference algorithms originally developed for the EP prior. Through simulations, we show that the M-EP prior combined  with VI achieves performance in ER tasks comparable to the methods of \citet{Bet22}, while reducing computation time by two orders of magnitude. Furthermore,  by leveraging stochastic VI (SVI; \citealp{hoffman2013stochastic}), we achieve an additional reduction in computational cost by an order of magnitude or more. Overall, our methods scale to datasets with tens of thousands of records in seconds or minutes on a standard laptop, making large-scale ER practically feasible.


\section{The microclustering Ewens--Pitman prior}

\subsection{The Ewens--Pitman prior}

The EP model \citep{Pitman95} is a two-parameter generalization of the celebrated Ewens model for random partitions \citep{Ewe72}. For $n\geq1$ let $\Pi_{n}$ be a random partition of the set $[n]=\{1,\ldots,n\}$ into $K_{n}\leq n$ blocks of sizes $(N_{1,n},\ldots,N_{K_{n},n})$, such that $N_{i,n}>0$ and $\sum_{1\leq i\leq K_{n}}N_{i,n}=n$. For $\alpha\in[0,1)$ and $\theta>0$, the EP model assigns to $\Pi_{n}$ the probability 
\begin{equation}\label{eppf_pit}
\text{Pr}[K_{n}=k,\,(N_{1,n},\ldots,N_{K_{n},n})=(n_{1},\ldots,n_{k})]\propto\frac{1}{k!}\prod_{i=1}^{k}\frac{(\theta+(i-1)\alpha)(1-\alpha)_{(n_{i}-1)}}{n_{i}!},
\end{equation}
where $(a)_{(u)}$ is the $u$-th rising factorial of $a>0$, i.e. $(a)_{(u)}=\prod_{0\leq i\leq u-1}(a+i)$ with the proviso $(a)_{(0)}=1$. We denote by $\Pi_{n}\sim\text{EP}(\alpha,\theta)$ the EP random partition; the case $\alpha=0$ corresponds to the Ewens random partition \citep[Chapter 2 and Chapter 3]{Pitman2006}. In the context of ER, and more broadly clustering tasks, \eqref{eppf_pit} is used as a prior distribution for the latent partition of data into clusters. Therefore, $K_n$ is the number of clusters and the $N_{i,n}$'s are the cluster's sizes.

The random partition $\Pi_{n}\sim\text{EP}(\alpha,\theta)$ is finite exchangeable, namely: for any fixed $n\geq 1$, the distribution \eqref{eppf_pit} is a symmetric function of the block's sizes $n_{i}$'s. Moreover, the sequence $\Pi=(\Pi_{n})_{n\geq1}$ defines an infinite exchangeable random partition (or exchangeable random partition of $\mathbb{N}$). This infinite random partition follows from the consistency property that the restriction to $[m]$ of $\Pi_{n}$ has the same distribution as $\Pi_{n}$, almost surely for all $m<n$. As a result, the distribution of $\Pi$ is invariant under all finite permutations of its elements \citep[Chapter 2]{Pitman2006}.

\subsection{Scaling the EP prior}

The M-EP prior is defined as a scaling, with respect to the sample size $n\geq1$, of the EP prior \eqref{eppf_pit}.

\begin{definition}\label{def_mep}
For $n\geq1$, the M-EP prior assigns to the random partition $\Pi_{n}$ of $[n]$ the probability \eqref{eppf_pit} with $\alpha\in[0,1)$ and $\theta=\lambda n$, for $\lambda>0$. We write $\Pi_{n}\sim\text{M-EP}(\alpha,\lambda )$ for the M-EP random partition.
\end{definition}

For any fixed $n\geq1$, the random partition $\Pi_{n}\sim\text{M-EP}(\alpha,\lambda )$ is finite exchangeable. Indeed, replacing $\theta$ with $\lambda n$ in \eqref{eppf_pit} preserves the symmetry of the distribution in the block's sizes $n_{i}$'s. However, the sequence $\Pi=(\Pi_{n})_{n\geq1}$ no longer defines an infinite exchangeable random partition since the scaling $\theta=\lambda n$ breaks the consistency property the the restrictions to $[m]$ of $\Pi_{n}$.

Let $N_{(1),n}$ be the largest block's size of $\Pi_{n}\sim\text{M-EP}(\alpha,\lambda)$. The next theorem shows that $N_{(1),n}$ grows  sub-linearly with $n$.

\begin{theorem}\label{teo:micro}
For $n\geq1$, $\alpha\in[0,1)$ and $\lambda>0$ let $\Pi_{n}\sim\text{M-EP}(\alpha,\lambda)$. Then, $n^{-1}N_{(1),n}\stackrel{\text{p}}{\longrightarrow}0$ as $n\rightarrow+\infty$.
\end{theorem}

See \ref{proof_micro} for the proof of Theorem \ref{teo:micro}. The next proposition shows that the number $K_{n}$ of blocks of $\Pi_{n}\sim\text{M-EP}(\alpha,\lambda)$, as well as the number $M_{r,n}$ of blocks of any fixed size $r\geq1$, grow linearly with $n$.

\begin{proposition}\label{teo:asym_k_m}
For $n\geq1$, $\alpha\in[0,1)$ and $\lambda>0$ let $\Pi_{n}\sim\text{M-EP}(\alpha,\lambda)$. The following holds true:
\begin{itemize}
\item[i)] if
 \begin{displaymath}
 \mathcal{M}_{\alpha,\lambda}:=\begin{cases} \lambda \log \left(\frac{\lambda+1}{\lambda}\right) &  \text{ for } \alpha = 0\\[0.2cm]
 \frac{\lambda}{\alpha}\left[\left(\frac{\lambda+1}{\lambda}\right)^{\alpha}-1\right]& \text{ for } \alpha \in (0, 1),
 \end{cases} 
 \end{displaymath}
then as $n\rightarrow+\infty$
\begin{equation}\label{mom_m_lln}
E\left[K_n\right] =n\mathcal{M}_{\alpha, \lambda} + O(1) \quad \text{and} \quad \frac{K_{n}}{n}\stackrel{\text{p}}{\longrightarrow}\mathcal{M}_{\alpha, \lambda};
\end{equation}
\item[ii)] for fixed $r\geq1$, if
 \begin{displaymath}
 \mathcal{M}_{\alpha,\lambda}(r):=\begin{cases} \frac{1}{r}\lambda(\lambda+1)^{-r} &  \text{ for } \alpha = 0\\[0.2cm]
 \frac{(1-\alpha)_{(r-1)}}{r!}\lambda^{1-\alpha}(\lambda+1)^{\alpha-r} & \text{ for } \alpha \in (0, 1),
 \end{cases} 
 \end{displaymath}
then as $n\rightarrow+\infty$
\begin{equation}\label{mom_m_r_llnr}
E\left[M_{r,n}\right] =n\mathcal{M}_{\alpha, \lambda}(r) + O(1) \quad \text{and} \quad \frac{M_{r,n}}{n}\stackrel{\text{p}}{\longrightarrow}\mathcal{M}_{\alpha, \lambda}(r).
\end{equation}
\end{itemize}
\end{proposition}

See \ref{proof_micro_blocks} for the proof of Proposition \ref{teo:asym_k_m}. The asymptotic behaviour of $K_{n}$ in \eqref{mom_m_lln} was established by \citet[Theorem 1]{Contardi24}, and it is reported in Proposition \ref{teo:asym_k_m} for completeness. Together, Theorem \ref{teo:micro} and Proposition \ref{teo:asym_k_m}, show the microclustering property of the M-EP prior.


\section{Variational inference algorithms for entity resolution}

\subsection{Entity resolution}\label{sec:er}

Let $X = (x_{i, \ell}, i=1, \ldots, n, \ell=1, \ldots, L)$ be the data matrix such that $x_{i, \ell} \in \{1, \ldots, D_\ell\}$ is the $\ell$-th attribute for the $i$-th sample. Following \cite{Bet22}, we consider a set of entities $(y_k, k \geq 1)$ with $y_{k} = (y_{k, 1}, \ldots, y_{k, L})$, such that $\text{Pr}[y_{k, \ell} = m] = \theta_{\ell, m}$, independently across $k$ and $\ell$, for $\theta_\ell = (\theta_{\ell, 1}, \ldots, \theta_{\ell, D_{\ell}})$ a probability vector. We assume that the data are generated as follows. First, the unique entities $(y_k)_{k \geq 1}$ are generated as above, together with the partition $\Pi_{n}\sim\text{M-EP}(\alpha,\lambda )$ of $[n]$. All data whose indices $i$ are in the $k$-th block of the partition are given a noise-free record $\tilde x_i = y_k$. Datum $x_i$ is a possibly noisy representation of $\tilde x_i$: let $\beta_\ell \in [0, 1)$ be the rate of distortion for the $\ell$-th feature, we assume that, with probability $(1 - \beta_\ell)$, $x_{i, \ell} = \tilde x_{i, \ell}$, while, with probability $\beta_\ell$, $x_{i, \ell} \sim \mathrm{Categorical}(\theta_\ell)$ independently across $i$ and $\ell$.

\subsection{Linking to the Pitman--Yor process}

The Pitman--Yor process \citep[PYP,][]{PitmanYor97} is a discrete random probability measure central to Bayesian nonparametrics, and it admits the following stick-breaking representation. For $\alpha\in[0,1)$ and $\theta>0$ let $\nu_{j}\ind\text{Beta}(1-\alpha,\theta+j\alpha)$, for $j\geq1$, and let $(y_{k})_{k\geq1}$ be i.i.d. random variables with non-atomic distribution $G_0$ on a measurable space $\mathbb{S}$, and independent of the $\nu_{j}$'s. By defining $p_{1}=\nu_{1}$ and $p_{k}=\nu_{k}\prod_{1\leq j\leq k-1}(1-\nu_{j})$ for $k\geq1$, such that $p_{k}\in(0,1)$ for all $k\geq1$ and $\sum_{k\geq1}p_{k}=1$ almost surely, the random probability measure $P=\sum_{k\geq1}p_{k}\delta_{y_{k}}$ on $\mathbb{S}$ is a PYP with strength $\theta$ and discount $\alpha$. The Dirichlet process corresponds to $\alpha=0$ \citep{Ferguson73,Sethuraman1994}. We write $P \sim \mathrm{PYP}(\alpha,\theta)$, omitting explicit reference to $G_{0}$, which plays no essential role in what follows.

Consider a random sample $x_i$ from $P \sim \mathrm{PYP}(\alpha,\theta)$, for $i=1, \ldots, n$. From \citet[Propostion 9]{Pitman95}, the random partition of $[n]$ induced by the equivalence relation $i \sim j$ if $x_i = x_j$ is distributed according to  the EP prior \eqref{eppf_pit}. Consequently, the generative scheme in Section \ref{sec:er} is equivalent to sampling $x_i$ conditionally i.i.d. from $P = \sum_{k \geq 1} p_k \delta_{y_k} \sim \mathrm{PYP}(\alpha,\lambda n)$, which is supported on all the latent entities, and then perturbing the noise-free records as above. Such a connection with the PYP will serve as a foundation for the VI schemes developed hereafter.

\subsection{Variational inference algorithm}

As in \cite{Bet22}, we treat $\theta_\ell$ and $\beta_\ell$ as fixed constants, although priors could be introduced. We approximate the posterior with variational inference (VI), which selects the approximate posterior $q(\cdot) \in \mathcal{Q}$ by minimising the Kullback–Leibler divergence, or equivalently maximising the evidence lower bound (ELBO). We adopt a  mean-field family of and retain only the first $K$ sticks in the stick-breaking representation of the PYP \citep{Blei06}; unlike \cite{ishwaran2001gibbs}, truncation is applied solely to $q$ (and not to the prior). The VI family is parametrized as follows:
\begin{equation}\label{eq:vi_full}
q(\bm v, \bm z, \bm y) = \prod_{k=1}^{K-1} q(v_k) \prod_{i=1}^n q(z_i) \prod_{k=1}^K \prod_{\ell=1}^L q(y_{k\ell}),
\end{equation}
where $q(v_k) = \mathrm{Beta}(v_k; a_k, b_k)$, $q(z_i) = \mathrm{Categorical}(z_i; r_{i1}, \dots, r_{iK})$, and $q(y_{k\ell}^*) = \mathrm{Categorical}(y_{k\ell}; \phi_{k\ell1}, \dots, \phi_{k_\ell D_\ell})$. Then, we optimize with respect to the variational parameters $(a_k, b_k)$, $r_{ik}$, and $\phi_{k\ell d}$ via coordinate ascent; see \ref{app:vi_full_family} for details. This algorithm allows to scale  to tens of thousands of datapoints in a matter of a few minutes, resulting in an average speed-up from the MCMC in \cite{Bet22} at least two orders of magnitude.
However, the mean-field ELBO exhibits many local optima and often yields sub-optimal estimates.

To improve accuracy we marginalise \(y_{k\ell}\), obtaining the collapsed family  $q(\bm v, \bm z) = \prod_{1\leq k\leq K-1} q(v_k) \prod_{1\leq i\leq n} q(z_i)$ and ELBO
\begin{equation}
  \mathcal L(q)=
     \E_q[\log p(\bm X\mid\bm z)]
     +\E_q[\log p(\bm z\mid\bm v)]
     +\E_q[\log p(\bm v)]
     -\E_q[\log q(\bm z)]
     -\E_q[\log q(\bm v)],
     \label{eq:full_elbo}
\end{equation}
where the term $ p(\bm X\mid\bm z)$ is the likelihood with the $y_k$'s marginalized out: 
\begin{displaymath}
  p(\bm X\mid\bm z)
  =  \prod_{\ell=1}^L \prod_{i=1}^n \beta_\ell \theta_{\ell x_{i \ell}} \times 
  \prod_{k=1}^{K}\prod_{\ell=1}^{L}
    f(\bm x_{k\ell}\mid\bm z),
\end{displaymath}
where $\bm x_{k\ell} = (x_{i\ell} \colon z_i = k)$, and
\begin{displaymath}
    f(\bm x_{k\ell}\mid\bm z) = 1-\sum_{d\in U_{k\ell}}\theta_{\ell d}
    +\sum_{d\in U_{k\ell}}
       \theta_{\ell d}
       \Bigl(1+\tfrac{1-\beta_\ell}{\beta_\ell\theta_{\ell d}}\Bigr)^{n_{k\ell d}},
\end{displaymath}
with $n_{k\ell d}=\sum_{1\leq i\leq n}\mathbf 1\{z_i=k,x_{i\ell}=d\}$, and $U_{k\ell}=\{d:n_{k\ell d}>0\}$. See \cite{Bet22} for details.

All terms in \eqref{eq:full_elbo} except the first one are straightforward, and follow from \cite{Blei06}. The first one splits into the constant
$\sum_{i\ell}\log(\beta_\ell\theta_{\ell,x_{i\ell}})$ plus $\E_q[\log f(\bm x_{k\ell}\mid\bm z)]$, which is intractable as it depends on the latent indicators $z_i$'s. 
To circumvent this obstacle, we invoke Jensen’s inequality, i.e.,
\begin{align*}
     \E_q\bigl[\log f(\bm x_{k\ell}\mid\bm z)\bigr]
  & \ge
  \log \E_q\bigl[f(\bm x_{k\ell}\mid\bm z)\bigr] \\
  & \geq  \log \Bigl[
        1
        +\sum_{d=1}^{D_\ell}
           \theta_{\ell d}
           \Bigl\{
              \Bigl(1+\tfrac{1-\beta_\ell}{\beta_\ell\theta_{\ell d}}\Bigr)^{\tilde n_{k\ell d}} - 1 
           \Bigr\}
      \Bigr] =: f_{\text{soft}}(\bm x_{k\ell}),
\end{align*}
where the $\tilde n_{k\ell d}= \sum_{1\leq i\leq n} r_{ik}\,\mathbf 1\{x_{i\ell}=d\}$. The bound now depends only on the ``soft counts'' $\tilde n_{k\ell d}$ and is therefore tractable. Using $f_{\text{soft}}$ in place of $ \E_q[\log p(\bm X\mid\bm z)]$ in \eqref{eq:full_elbo} leads to a new objective function $\tilde{\mathcal L}(q)$ that  is still a lower bound on the evidence.

With cached quantities, optimizing $\tilde{\mathcal L}(q)$ nearly as fast as the naive VI but markedly more accurate for entity resolution; see \ref{app:vi_collapsed_family} for detail, and  Algorithm \ref{algo:cvi} for the pseudocode. In practice, we assume suitable hyper-priors for the PYP parameters as well, i.e., $\lambda \sim \mathrm{Beta}(a_\lambda, b_{\lambda})$ and $\alpha \sim \mathrm{Beta}(a_\alpha, b_\alpha)$, see Appenidix \ref{app:hyperparams} for the corresponding updates in the VI algorithm.

\begin{algorithm}[H]
\caption{Collapsed VI algorithm}
\label{algo:cvi}
\Repeat{ELBO converges}{
  
    Let $\gamma_k\leftarrow \psi(a_k)-\psi(a_k+b_k)
                       +\sum_{j<k}\bigl[\psi(b_j)-\psi(a_j+b_j)\bigr]$

    Compute soft-counts $\tilde n_{k \ell d}$

    Let $\log r_{ik}\propto
  \gamma_k
  +\sum_{\ell=1}^{L}
      \log\frac{f_{\text{soft}}\bigl(\bm x_{k\ell}^{+i}\bigr)}
                     {f_{\text{soft}}\bigl(\bm x_{k\ell}^{-i}\bigr)},$
    where $\bm x_{k\ell}^{+i}$ (\emph{resp.} $-i$) is the count vector
with record $i$ included (resp. excluded);

  \BlankLine
  
  \For{$k\gets1$ \KwTo $K$}{
        $a_k\leftarrow 1-\alpha+N_k$ for $N_k\leftarrow\sum_i r_{ik}$
        $b_k\leftarrow \theta+\alpha(k-1)+\sum_{j>k}N_j$ }

}
\end{algorithm}

\subsection{Stochastic variational inference and further algorithmic improvements}\label{sec:svi}

The bottleneck of the collapsed VI algorithm is the update of the variational parameters $r_{ik}$, which require updating an $n \times K$ matrix at every iteration. Since in microclustering the number of clusters scales linearly with $n$, this means that the update is $O(n^2)$. There are tweaks to mitigate the computational burden. First, is the use of stochastic VI \citep[SVI,][]{hoffman2013stochastic}, whereby at each iteration, only a mini-back of $m$ datapoints is considered, and the corresponding $r_i$'s updated.  Since $m \ll n$, this leads to substantial speed-ups. See \ref{app:svi} for the details.

In very large datasets, or when memory budget is constrained, storing the full responsibility matrix $r_{ik}$ might be unfeasible. This is because the truncation level $K$ needs to scale linearly with $n$, thus requiring $O(n^2)$ memory. However, especially in microclusterin tasks, it is to be expected that all but a few of the $r_{ik}$'s are essentially zero. This makes our setting a perfect candidate for the sparse posteriors of \cite{hughes2016fast}, whereby for each $i$, at most $V$ $r_{ik}$'s are allowed to be different from zeros. We refer to this tweak as the top-$V$ thresholding.
Using sparse matrix algebra, this leads to substantial memory improvements. See \cite{hughes2016fast} for further details.


\section{Numerical illustrations}

\subsection{Synthetic data generation}\label{sec:data}

For fixed $n, L$ and $D_\ell$ (to be specified later), we let $\theta_{\ell, j} = 1/D_{\ell}$ for any $\ell, j$, and let $\beta_\ell \in \{0.01, 0.05\}$.
We generate data by first sampling $M$ latent entities $y_{k}$ iid from the product of categorical distributions with parameters $\theta_\ell$, and the true cluster allocations $z_i$ from the discrete uniform over $\{1, \ldots, M\}$. On average, we expect that only $K_n < M$ entities are selected. Letting $\tilde{x}_{i} = y_{z_i}$, we obtain the data $x_i$ by perturbing the noise-free records as described in Section \ref{sec:er}.

\subsection{Benchmarking against \cite{Bet22}}\label{sec:simu1}

We compare inference obtained with Algorithm \ref{algo:cvi} against the model of \cite{Bet22} (specifically, we consider here only their ``ESCD'' prior, as it is the one that performs better in practice). We compute the adjusted rand index (ARI) between the true and estimated partitions. For our model, we obtain a point estimate of the partition by taking the arg-max (row-wise) of the $r_{ik}$'s. For the ESCD model, we compute the average ARI from the MCMC output. Since the runtime of the MCMC method in \cite{Bet22} is potentially unbounded, we consider two approaches. The first one runs the MCMC for 2000 iterations, discarding the first 500 iterations as burn-in, as suggested in \cite{Bet22}.The second one caps the number of iterations to a much smaller value in order to have the same runtime as our VI algorithm on average. For the dataset size considered here, running 2000 MCMC iterations requires approximately 10-12 minutes, while fitting the VI algorithm takes between 5 to 10 seconds, i.e., an 85x speed-up.

We generate data as in Section \ref{sec:data} with $L=5$, $n=2000$, $D_\ell = 10$ for every $\ell$, and $M=500$. The VI algorithm is fitted with truncation $K=2M$.
A priori, we assume that $\alpha$ and $\lambda$ are Beta distributed with parameters $(2, 2)$.
Table \ref{tab:simu1} summarizes our findings over 50 independent replicates. When $\beta=0.01$, VI seems to perform as good as the MCMC with full iterations, while the performance slightly degrades when $\beta=0.05$. However, VI clearly outperforms the short version of the MCMC, showing that the VI algorithm offers a feasible alternative to full MCMC when the dataset sizes become impractical.

\begin{table}
\centering
\caption{Mean, 5\%, and 95\% quantile of the ARI between estimated and true partition for the simulation in Section \ref{sec:simu1}. \vspace{1em}}
\label{tab:simu1}
\begin{tabular}{l|ccc|ccc|ccc}
\toprule
Method & \multicolumn{3}{c|}{VI} & \multicolumn{3}{c}{MCMC (short)} & \multicolumn{3}{|c}{MCMC (full)} \\
 & mean & $q_{0,05}$ & $q_{0,95}$ & mean & $q_{0,05}$ & $q_{0,95}$& mean & $q_{0,05}$ & $q_{0,95}$ \\
\hline
$\beta=0.01$ & 0.96 & 0.95 & 0.97 & 0.63 & 0.56 & 0.70 & 0.97 & 0.96 & 0.97 \\
$\beta=0.05$ & 0.86 & 0.84 & 0.88 & 0.56 & 0.52 & 0.62 & 0.89 & 0.88 & 0.90 \\
\end{tabular}
\end{table}

\subsection{Scalability of the VI algorithm}

We further explore the scalability of the VI algorithm and the variants described in Section \ref{sec:svi}. Data are generated as in Section \ref{sec:data}, letting $n\in\{5000, 10000, 15000, 20000\}$ with $D_\ell = 5, 7, 9,  10$ respectively, $M=N/4$ and truncation $K=2M$. Here we assume $\alpha=0.25$ and $\lambda=0.5$ are fixed for all the models. Even for the smallest setting considered here, the runtime of the full MCMC would require more than 10 hours making its application unfeasible.

We compare the collapsed VI algorithm with its stochastic counterpart, in which we optionally include the top-$V$ thresholding of \cite{hughes2016fast} for $V \in \{8, 16, 32\}$.
Table \ref{tab:simu2} reports out findings, averaged over 50 independent replicates. For $n=20000$, we excluded th ``full'' VI algorithm due to excessive runtimes. From Table \ref{tab:simu2}, it is clear that the stochastic approximations of the full VI algorithm provide reliable estimates, often outpeforming slightly the full VI. This is  in accordance with the fact that stochastic optimization is able to excape local extrema more easily \citep{Klei18}, therefore reaching better global solutions.
The SVI algorithm reduces computations by a factor of around 6--17 (when $n=5000$ and $n=15000$, respectively) and is the fastest across all settings, thanks to the use of vectorized operations. On the other hand, the top-$K$ thresholding reduces memory usage by a factor of 2 in all settings compared to SVI.

\begin{table}
\centering
\caption{Performance summary by $n$, $L$, $\beta$ and engine. \vspace{1em}}
\label{tab:simu2}
\small
\begin{tabular}{c c c | c c c c c}
\toprule
 $n$ & $\beta$ & metric   & Full-VI & SVI & SVI-V:16 & SVI-V:32 & SVI-V:64 \\
\midrule
\multirow[c]{4}{*}{5000} & \multirow[c]{2}{*}{0.01} & ARI & 0.93 & \textbf{0.94} & 0.79 & 0.85 & 0.88 \\
 &  & Time & 223.61 & \textbf{35.86} & 87.57 & 73.48 & 54.38 \\
 \cline{2-8}
& \multirow[c]{2}{*}{0.05} & ARI & \textbf{0.75} & 0.71 & 0.45 & 0.58 & 0.67 \\
 &  & Time & 127.52 & \textbf{37.81} & 89.25 & 89.39 & 87.28 \\
\cline{1-8} \cline{2-8}
\multirow[c]{4}{*}{10000} & \multirow[c]{2}{*}{0.01} & ARI & 0.83 & 0.96 & 0.86 & 0.94 & \textbf{0.97} \\
 &  & Time & 591.01 & \textbf{98.41} & 154.61 & 152.15 & 123.43 \\
\cline{2-8}
 & \multirow[c]{2}{*}{0.05} & ARI & \textbf{0.87} & 0.82 & 0.51 & 0.68 & 0.80 \\
 &  & Time & 449.93 & \textbf{96.27} & 153.44 & 154.55 & 154.70 \\
\cline{1-8} \cline{2-8}
\multirow[c]{4}{*}{15000} & \multirow[c]{2}{*}{0.01} & ARI & 0.66 & 0.96 & 0.87 & 0.95 & \textbf{0.97} \\
 &  & Time & 1347.62 & \textbf{178.10} & 234.95 & 235.91 & 218.54 \\
\cline{2-8}
 & \multirow[c]{2}{*}{0.05} & ARI & \textbf{0.90} & 0.78 & 0.47 & 0.69 & 0.84 \\
 &  & Time & 1517.67 & \textbf{185.06} & 238.63 & 238.17 & 241.28 \\
\cline{1-8} \cline{2-8}
\multirow[c]{4}{*}{20000} & \multirow[c]{2}{*}{0.01} & ARI & -- & 0.96 & 0.81 & 0.93 & \textbf{0.97} \\
 &  & Time & -- & \textbf{265.05} & 311.22 & 309.80 & 314.36 \\
\cline{2-8}
 & \multirow[c]{2}{*}{0.05} & ARI & -- & 0.72 & 0.48 & 0.69 & \textbf{0.84} \\
 &  & Time & -- & \textbf{267.88} & 320.56 & 310.82 & 313.08 \\
\bottomrule
\end{tabular}
\end{table}


\section{Discussion}

By enabling Bayesian inference for ER in the large $n$ setting, the M-EP prior brings frequentist properties back into focus for practitioners. \cite{johndrow2018theoretical} present an impossibility result for ER as $n\rightarrow+\infty$, but \cite{Bet22} argue this can be overcome if the attribute dimension $L$ grows with $n$. Still, consistency in ER is a nonstandard statistical problem. It is natural to assume data are generated entity-wise: a latent entity is drawn from a population distribution, then noisy replicas are produced. Proving (or refuting) posterior consistency in this mixture-of-mixtures setting will require new tools.

\bibliographystyle{apalike}
\bibliography{references}

\clearpage

\section*{Supplementary material to: Large-scale entity resolution via microclustering Ewens--Pitman random partitions}\label{SM}

\appendix
 \setcounter{page}{1}

\numberwithin{equation}{section}
\numberwithin{theorem}{section}
\numberwithin{lemma}{section}


\section{Proofs}

\subsection{Proof of Theorem \ref{teo:micro}}\label{proof_micro}

We treat separately the case $\alpha=0$ and the case $\alpha\in(0,1)$, with $\lambda>0$. We start with the case $\alpha=0$. From \citet{Kin(78),Kin(82),Ald(85)} and \citet[Corollary 18]{Pit(97)}, for any $r\geq1$ as $n\rightarrow+\infty$
\begin{equation}\label{e1_case0}
\E\left[\left(\frac{N_{(1),n}}{n}\right)^{r}\right]\approx\frac{\Gamma(\lambda n+1)}{\Gamma(\lambda n+r)}\int_{0}^{+\infty}t^{r-1}\text{e}^{-t-\lambda n E(t)}\ddr t,
\end{equation}
where
\begin{displaymath}
E(t)=\int_{t}^{+\infty}\frac{1}{x}\text{e}^{-x}\ddr x.
\end{displaymath}
See also \citet[Section 2.4 and Chapter 4]{Pit(06)} and references therein for details on \eqref{e1_case0}. We apply Laplace's method on order to obtain a large $n$ approximation of the integral on the right-hand side of \eqref{e1_case0}, i.e.  
\begin{displaymath}
I_{n}=\int_{0}^{+\infty}t^{r-1}\text{e}^{-t-\lambda n E(t)}\ddr t,
\end{displaymath}
where we set $f_n(t)= t^{r-1} \mathrm{e}^{-\phi_{n}(t)}$ with $\phi_{n}(t)= t + \lambda n E(t)$. By taking the (first) derivative $\phi_{n}'(t) = 1 - \lambda n t^{-1}e^{-t}$ and setting such a derivative equal to $0$, we obtain an implicit equation for the saddle point $t_{n}$. That is,
\begin{equation}\label{eq:sadd}
t_{n}\text{e}^{t_{n}}=\lambda n,
\end{equation}
which can not be solved explicitly in $t_{n}$. However, since $\lambda n\geq-\text{e}^{-1}$, \eqref{eq:sadd} admits a solution in terms of a Lambert function \citep[Section 4.13]{Olv(10)}. Denoting by $W$ the Lambert function, $t_{n}=W(\lambda n)$ such that, as $n\rightarrow+\infty$
\begin{displaymath}
t_n \approx \log(\lambda n) - \log\log(\lambda n)>0.
\end{displaymath}
Now, consider the second derivative
\begin{displaymath}
\phi_{n}''(t) = \lambda n \left( \frac{e^{-t}}{t} +\frac{e^{-t}}{t^2} \right) = \lambda n \frac{e^{-t}(t + 1)}{t^2}.
\end{displaymath}
such that, as $t\rightarrow+\infty$
\begin{displaymath}
\phi_{n}''(t) = \frac{t + 1}{t}\approx 1.
\end{displaymath}
Since $E(t)\approx t^{-1}\text{e}^{-t}$ as $t\rightarrow+\infty$, then $E(t_{n})\approx(\lambda n)^{-1}$, as $n\rightarrow+\infty$. The leading behavior of the integrand at its peak is 
\begin{displaymath}
f_n(t_{n}) = t_{n}^{r-1} \mathrm{e}^{-t_{n} - 1}
\end{displaymath}
such that, as $n\rightarrow+\infty$
\begin{displaymath}
I_{n} \approx t_{n}^{r-1} \mathrm{e}^{-t_{n}- 1}\sqrt{\frac{2\pi}{|\phi_{n}''(t_{n})|}}\approx t_{n}^{r-1} \mathrm{e}^{-t_{n} - 1}\sqrt{2\pi}.
\end{displaymath}
From the asymptotic behaviour of the ratio of Gamma functions \citep[Equation 1]{Tri(51)}, as $n\rightarrow+\infty$
\begin{displaymath}
\frac{\Gamma(\lambda n+1)}{\Gamma(\lambda n+r)}\int_{0}^{+\infty}t^{r-1}\text{e}^{-t-\lambda n E(t)}\ddr t\approx(\lambda n)^{-r}t_{n}^{r-1} \mathrm{e}^{-t_{n} - 1}\sqrt{2\pi},
\end{displaymath}
where, as $n\rightarrow+\infty$
\begin{displaymath}
t_{n}^{r-1}\approx (\log n)^{r-1}
\end{displaymath}
and
\begin{displaymath}
 \mathrm{e}^{-t_{n}} \approx \frac{1}{\lambda n}.
\end{displaymath}
Hence, as $n\rightarrow+\infty$
\begin{displaymath}
\E\left[\left(\frac{N_{(1),n}}{n}\right)^{r}\right]\approx\frac{\Gamma(\lambda n+1)}{\Gamma(\lambda n+r)}\int_{0}^{+\infty}t^{r-1}\text{e}^{-t-\lambda n E(t)}\ddr t\approx\frac{\sqrt{2\pi}}{e}\frac{(\log n)^{r - 1}}{(\lambda n)^r}\rightarrow0.
\end{displaymath}
Since $\E[n^{-1}N_{(1),n}]$ and $\E[n^{-2}N^{2}_{(1),n}]$ go to $0$ as $n\rightarrow+\infty$, the proof is completed by an application of Chebyshev inequality.

Now, consider the case $\alpha\in(0,1)$, which is along lines similar to the case $\alpha=0$, though with differences. From \citet{Kin(78),Kin(82),Ald(85)} and \citet[Proposition 17]{Pit(97)}, for any $r\geq1$ as $n\rightarrow+\infty$
\begin{equation}\label{e1_case1}
\E\left[\left(\frac{N_{(1),n}}{n}\right)^{r}\right]\approx(\Gamma(1-\alpha))^{\lambda n/\alpha}\frac{\Gamma(\lambda n+1)}{\Gamma(\lambda n +r)}\int_{0}^{+\infty}t^{r+\lambda n-1}\text{e}^{-t}(F(t))^{-1-\lambda n/\alpha}\ddr t,
\end{equation}
where
\begin{displaymath}
F(t)=\Gamma(1-\alpha)t^{\alpha}+\alpha\int_{1}^{+\infty}\text{e}^{-tx}x^{-\alpha-1}\ddr x.
\end{displaymath}
See also \citet[Section 2.4 and Chapter 4]{Pit(06)} and references therein for details on \eqref{e1_case1}. We apply Laplace's method in order to obtain a large $n$ approximation of the integral on the right-hand side of \eqref{e1_case1}, i.e.  
\begin{displaymath}
I_n =\int_{0}^{+\infty}\text{e}^{\phi_{n}(t)}\ddr t,
\end{displaymath}
where
\begin{displaymath}
\phi_{n}(t) = (r+\lambda n-1)\log(t)-t-\left(1+\frac{\lambda n}{\alpha}\right)\log(F(t)).
\end{displaymath}
We consider $r\geq2$; the case $r=1$ then follows by a direct application of Holder inequality. As  $n\rightarrow+\infty$, $\phi_{n}(t)$ as a maximum for $t\rightarrow+\infty$; that is, we are interested in large values of $t$. To find the critical point of $\phi_{n}(t)$, let
\begin{displaymath}
\phi_{n}^{'}(t)=\frac{r+\lambda n-1}{t}-1-\left(1+\frac{\lambda n}{\alpha}\right)\frac{F^{'}(t)}{F(t)}
\end{displaymath}
and set such a derivative equal to $0$. Then, we obtain an implicit equation for the saddle point $t_{n}$ (as in the case $\alpha=0$). That is,
\begin{displaymath}
\frac{r+\lambda n-1}{t_{n}}-1=\left(1+\frac{\lambda n}{\alpha}\right)\frac{F^{'}(t_{n})}{F(t_{n})}.
\end{displaymath}
Since $F(t)\approx\Gamma(1-\alpha)t^{\alpha}$ as $t\rightarrow+\infty$, as well as the derivative $F^{'}(t)\approx\Gamma(1-\alpha)\alpha t^{\alpha-1}$ as $t\rightarrow+\infty$, we have that,   as $t\rightarrow+\infty$,
\begin{displaymath}
\frac{r+\lambda n-1}{t_{n}}-1\approx\left(1+\frac{\lambda n}{\alpha}\right)\frac{\alpha}{t_{n}}.
\end{displaymath}
\begin{remark}
The saddle point $t_{n}$ increases in $n$, that is for $n^{'}<n^{''}$ it is expected $t_{n^{'}}\leq t_{n^{''}}$. However, as $t\rightarrow+\infty$,
\begin{displaymath}
\frac{r+\lambda n-1}{t_{n}}-1\approx\left(1+\frac{\lambda n}{\alpha}\right)\frac{\alpha}{t_{n}},
\end{displaymath}
such that $t_{n}\approx t_{0}=r-1-\alpha$. One could be more precise by considering a more precise asymptotics of $F(t)$ as $t\rightarrow+\infty$.
\end{remark}
Now, we proceed with the application of Laplace’s method. In particular, we consider the second derivative of $\phi_{n}(t)$, i.e.,
\begin{displaymath}
\phi_{n}''(t) =-\frac{r+\lambda n-1}{t^{2}}-\left(1+\frac{\lambda n}{\alpha}\right)\frac{F^{''}(t)F(t)-F^{'}(t)F^{'}(t)}{(F(t))^{2}}
\end{displaymath}
such that, as $t\rightarrow+\infty$
\begin{displaymath}
\phi_{r,n}''(t)\approx\frac{-r+1+\alpha}{t^{2}}.
\end{displaymath}
Therefore, by combining the above calculations to the large $n$ approximation of $I_{n}$, we can write that as $n\rightarrow+\infty$
\begin{displaymath}
I_{n}\approx\text{e}^{\phi_{n}(t_{n})}\sqrt{\frac{2\pi}{|\phi_{n}^{''}(t_{n})|}},
\end{displaymath}
where
\begin{align*}
\phi_{n}(t_{n})&\approx(r+\lambda n-1)\log(t_{0})-t_{0}-\left(1+\frac{\lambda n}{\alpha}\right)(\log(\Gamma(1-\alpha))+\alpha\log(t_{0}))=-\frac{\lambda n}{\alpha}\log(\Gamma(1-\alpha))+C_{0},
\end{align*}
with $C_{0}=r\log(t_{0})-\log(t_{0})-t_{0}-\log(\Gamma(1-\alpha))-\alpha\log(t_{0})$, that is $\text{e}^{C_{0}}=\Gamma(1-\alpha)t_{0}^{r-1-\alpha}\text{e}^{-t_{0}}$, and where
\begin{displaymath}
\phi_{r,n}''(t_{n})\approx\frac{1}{-t_{0}}.
\end{displaymath}
Then, as $n\rightarrow+\infty$
\begin{align*}
&(\Gamma(1-\alpha))^{\lambda n/\alpha}\frac{\Gamma(\lambda n+1)}{\Gamma(\lambda n +r)}\int_{0}^{+\infty}t^{r+\lambda n-1}\text{e}^{-t}(F(t))^{-1-\lambda n/\alpha}\ddr t.\\
&\quad=(\Gamma(1-\alpha))^{\lambda n/\alpha}\frac{\Gamma(\lambda n+1)}{\Gamma(\lambda n +r)}I_{n}\\
&\quad\approx(\Gamma(1-\alpha))^{\lambda n/\alpha}\frac{\Gamma(\lambda n+1)}{\Gamma(\lambda n +r)}\text{e}^{-\frac{\lambda n}{\alpha}\log(\Gamma(1-\alpha))}\text{e}^{C_{0}}\sqrt{\frac{2\pi}{\frac{1}{t_{0}}}}\\ 
&\quad=\frac{\Gamma(\lambda n+1)}{\Gamma(\lambda n +r)}\text{e}^{(\frac{\lambda n}{\alpha}-\frac{\lambda n}{\alpha})\log(\Gamma(1-\alpha))+C_{0}}\sqrt{\frac{2\pi}{\frac{1}{t_{0}}}}\\ 
&\quad=\frac{\Gamma(\lambda n+1)}{\Gamma(\lambda n +r)}\text{e}^{C_{0}}\sqrt{\frac{2\pi}{\frac{1}{t_{0}}}}\\
&\quad\approx n^{1-r}\text{e}^{C_{0}}\sqrt{\frac{2\pi}{\frac{1}{t_{0}}}}\\
&\quad\rightarrow0.
\end{align*}
The case $r=1$ follows from the case $r\geq2$. In particular, by considering $r\geq2$, from Holder inequality we have that
\begin{displaymath}
\E\left[\frac{N_{(1),n}}{n}\right]\leq\left(\E\left[\left(\frac{N_{(1),n}}{n}\right)^{r}\right]\right)^{1/r},
\end{displaymath}
where, as $n\rightarrow+\infty$
\begin{displaymath}
\left(\E\left[\left(\frac{N_{(1),n}}{n}\right)^{r}\right]\right)^{1/r}\approx n^{1/r-1}\left(\text{e}^{C_{0}}\sqrt{\frac{2\pi}{\frac{1}{t_{0}}}}\right)^{1/r}\\
\end{displaymath}
such that $E[n^{-1}N_{(1),n}]$ as $n\rightarrow+\infty$. Since $\E[n^{-1}N_{(1),n}]$ and $\E[n^{-2}N^{2}_{(1),n}]$ go to $0$ as $n\rightarrow+\infty$, the proof is completed by an application of Chebyshev inequality. This completes the proof for the whole range $\alpha\in[0,1)$ and $\lambda>0$.

\subsection{An alternative proof of Theorem \ref{teo:micro}}
We present an alternative proof of Theorem \ref{teo:micro}. The proof does not rely on \citet[Corollary 18]{Pit(97)} and \citet[Proposition 17]{Pit(97)}, allowing us to consider jointly the cases $\alpha=0$ and $\alpha\in(0,1)$. Let $N_{j, n}$, for $j=1, \ldots, K_n$ be the size of the $j$-th cluster, in order of appearance, in a random sample of size $n$ from $P\sim\mathrm{PYP}(\alpha,\lambda n)$. Moreover, denote by $N_{(1), n}$ the size the largest cluster. Here, we aim at showing 
\begin{equation}\label{eq:microcluster_def}
    \text{Pr}\left[\frac{N_{(1), n}}{n} > \varepsilon \right] \rightarrow 0
\end{equation}
for any $\varepsilon > 0$. Let $Y(\varepsilon) = \sum_{j \ge 1} I[N_{j, n} > n \varepsilon]$. Then, $\{N_{(1), n} > n \varepsilon\}$ is clearly contained in the event $\{Y(\varepsilon) \geq 1\}$.
Hence 
\[
    \text{Pr}[N_{(1), n} > n \varepsilon] \leq \text{Pr}\left[Y(\varepsilon) \geq 1\right] \leq E[Y(\varepsilon)]
\]
By exchangeability,
\begin{equation}\label{eq:exp_y}
   E[Y(\varepsilon)] = E\left[K_n \text{Pr}[N_{j, n} > \varepsilon n]\right].
\end{equation}
Conditionally to $P_j$ (the $j$-th weight in the size-biased representation of the PYP), $N_{j, n} \sim \operatorname{Binom}(n, P_j)$, such that
\begin{align}
    \text{Pr}\left[\frac{N_{j, n}}{n} > \varepsilon\right] &= \text{Pr}\left[\frac{N_{j, n}}{n} > \varepsilon, P_j > \delta \right] + \text{Pr}\left[\frac{N_{j, n}}{n} > \varepsilon, P_j \leq \delta \right] \nonumber \\
    & \leq \text{Pr}\left[P_j > \delta\right] + \text{Pr}\left[\frac{N_{j, n}}{n} > \varepsilon, P_j \leq \delta \right]. \label{eq:size_bound}
\end{align}
Now, we provide an upper bound for each of the terms in \eqref{eq:size_bound}, separately. In particular, according to the stick-breaking representation of the PYP, $P_j \leq V_j$ where $V_j \sim \operatorname{Beta}(1 - \alpha, \theta_n + j \alpha )$ where $\theta_n =\lambda n$. Therefore,
\[
    \text{Pr}\left[P_j > \delta\right) \leq \text{Pr}\left(V_j > \delta\right] = \frac{\mathrm B (1 - \delta; 1 - \alpha, \theta_n + j \alpha)}{\mathrm{B}(1 - \alpha, \theta_n + j \alpha)}
\]
where $\mathrm B(a, b)$ denotes the Beta function and $\mathrm B(x; a, b)$ denotes the incomplete bBeta function.
By the elementary bound
\[
    \int_\delta^1 t^{a - 1} (1 - t)^{b - 1} \mathrm d t \leq \frac{1}{b} \delta^{a - 1} (1 - \delta)^{b},
\]
we obtain
\[
   \text{Pr}\left[P_j > \delta\right] \leq \frac{\delta^{- \alpha}}{(\theta_n + j \alpha)} (1 - \delta)^{\theta_n + j \alpha} \leq C (1 - \delta)^{\lambda n},
\]
where $C$ does not depend on $n$. Consider now the second term in \eqref{eq:size_bound}, and bound it by Chernoff-Hoeffding inequality, i.e.,
\begin{align*}
    \text{Pr}\left[\frac{N_{j, n}}{n} > \varepsilon, P_j \leq \delta \right] = E\left[\indic[P_j \leq \delta] \mathrm P(N_{j, n} > \varepsilon n \mid P_j)\right],
\end{align*}
where
\[
     \text{Pr}[N_{j, n} > \varepsilon n \mid P_j] \leq e^{-n \mathrm{KL}(\varepsilon || P_j)}
\]
with
\[
    \mathrm{KL}(x || y) = x \log \frac{x}{y} + (1 - x) \log \frac{1 - x}{1-y}
\]
being the Kullback--Leibler divergence between two Bernoulli distributions with parameters $x$ and $y$, respectively. Since the divergence $\mathrm{KL}(x || y)$ is a decreasing function of $y$ when $y \in (0, x)$ we have for $P_j \leq \delta < \varepsilon$
\[
   \text{Pr}[N_{j, n} > \varepsilon n \mid P_j] \leq e^{-n \mathrm{KL}(\varepsilon || \delta)}.
\]
Therefore, choosing $\delta = \varepsilon/2$, 
\begin{displaymath}
\text{Pr}\left[\frac{N_{j, n}}{n} > \varepsilon, P_j \leq \delta \right] \leq e^{-n \mathrm{KL}(\varepsilon || \delta)} \leq e^{- 1/2 \varepsilon^2 n}.
\end{displaymath}
Hence,
\[
    \text{Pr}\left[\frac{N_{j, n}}{n} > \varepsilon\right] \leq C_{j, \alpha, \varepsilon} e^{-c_{\varepsilon, \lambda} n},
\]
where we have explicitly defined the dependence of the constants on the different parameters. Returning to \eqref{eq:exp_y},
\[
    E[Y(\varepsilon)] = E[K_N] \text{Pr}\left[\frac{N_{j, n}}{n} > \varepsilon\right] \leq C_{j, \alpha, \varepsilon, \lambda} \ n \ e^{-c_{\varepsilon, \lambda} n}
\]
which goes to zero as $n\rightarrow+\infty$, showing \eqref{eq:microcluster_def}. This completes the proof for the whole range $\alpha\in[0,1)$ and $\lambda>0$.

\subsection{Proof of Proposition \ref{teo:asym_k_m}}\label{proof_micro_blocks}

The proof of \eqref{mom_m_lln} was established by \citet[Theorem 1]{Contardi24}. Here, we present the proof of \eqref{mom_m_r_llnr}. We treat separately the cases $\alpha=0$ and $\alpha\in(0,1)$. We start with the case $\alpha=0$, for which we apply \citet[Proposition 1]{Fav(13)}; see also \citet[Section 3.1]{Fav(13)} for the case $\alpha=0$. In particular, we have
\begin{align*}
\E[M_{r,n}]&=\Gamma(r){n\choose r}\frac{(\lambda n)_{(n-r)}}{(\lambda n+1)_{(n-1)}}=\frac{1}{r}\frac{\Gamma(n+1)}{\Gamma(n-r+1)}\frac{\Gamma(n(\lambda+1)-r)\Gamma(\lambda n+1)}{\Gamma(\lambda n)\Gamma(n(\lambda +1))}.
\end{align*}
From the asymptotic behaviour of the ratio of Gamma functions \citep[Equation 1]{Tri(51)}, as $n\rightarrow+\infty$
\begin{equation}\label{eq:krmean0}
\E[M_{r,n}]=n\mathcal{M}_{\lambda}(r)+O(1),
\end{equation}
where
\begin{displaymath}
\mathcal{M}_{\lambda}(r)=\frac{1}{r}\lambda(\lambda+1)^{-r},
\end{displaymath}
which completes the proof of the large $n$ behaviour of $\E[M_{r,n}]$ in \eqref{mom_m_r_llnr}. Still from \citet[Proposition 1]{Fav(13)},
 \begin{align*}
\V[M_{r,n}]&=\left(\frac{\lambda n}{r}\right)^{2}(2r)!{n\choose 2r}\frac{(\lambda n)_{(n-2r)}}{(\lambda n)_{(n)}}\\
&\quad\quad+\Gamma(r){n\choose r}\frac{(\lambda n)_{(n-r)}}{(\lambda n+1)_{(n-1)}}\left(1-\Gamma(r){n\choose r}\frac{(\lambda n)_{(n-r)}}{(\lambda n+1)_{(n-1)}}\right)\\
&=\left(\frac{\lambda n}{r}\right)^{2}\frac{\Gamma(n+1)}{\Gamma(n-2r+1)}\frac{\Gamma(n(\lambda+1)-2r)\Gamma(\lambda n)}{\Gamma(\lambda n)\Gamma(n(\lambda+1))}\\
&\quad\quad+\frac{1}{r}\frac{\Gamma(n+1)}{\Gamma(n-r+1)}\frac{\Gamma(n(\lambda+1)-r)\Gamma(\lambda n+1)}{\Gamma(\lambda n)\Gamma(n(\lambda +1))}\\&\quad\quad-\left(\frac{1}{r}\frac{\Gamma(n+1)}{\Gamma(n-r+1)}\frac{\Gamma(n(\lambda+1)-r)\Gamma(\lambda n+1)}{\Gamma(\lambda n)\Gamma(n(\lambda +1))}\right)^{2}.
\end{align*}
From the asymptotic behaviour of the ratio of Gamma functions \citep[Equation 1]{Tri(51)}, as $n\rightarrow+\infty$
\begin{equation}\label{eq:krvar0}
\V[M_{r,n}]=n\mathcal{S}^{2}_{\lambda}(r)+O(1),
\end{equation}
where
\begin{align*}
\mathcal{S}^{2}_{\alpha,\lambda}(r)&=\frac{1}{r}(\lambda+1)^{-r}\lambda+\frac{1}{r^{2}}(\lambda+1)^{-2r}\lambda^{2}\left(\frac{\lambda^{2}r^{2}}{\lambda(\lambda+1)}\right).
\end{align*}
The proof of the law of large numbers in \eqref{mom_m_r_llnr} follows by an application of  \eqref{eq:krmean0} and \eqref{eq:krvar0}. In particular, we write
\begin{equation}\label{eq:decom0}
\frac{M_{r,n}-n\mathcal{M}_{\alpha,\lambda}(r)}{n}=\frac{M_{r,n}-\E[M_{r,n}]}{n}+\frac{\E[M_{r,n}]-n\mathcal{M}_{\lambda}(r)}{n},
\end{equation}
where:
\begin{itemize}
\item[i)] from \eqref{eq:krmean0}, as $n\rightarrow+\infty$
\begin{displaymath}
\frac{\E[M_{r,n}]-n\mathcal{M}_{\lambda}(r)}{n}\rightarrow0;
\end{displaymath}
\item[ii)] from \eqref{eq:krvar0}, for any $\varepsilon>0$ as $n\rightarrow+\infty$
\begin{displaymath}
\text{Pr}[|M_{r,n}-\E[M_{r,n}]|>n\varepsilon]\leq\frac{\V[M_{r,n}]}{n^{2}\varepsilon^{2}}=O(n^{-1}).
\end{displaymath}
\end{itemize}
Therefore, it holds $n^{-1}(M_{r,n}-\E[M_{r,n}])\stackrel{\text{p}}{\longrightarrow}0$ as $n\rightarrow+\infty$, which, according to \eqref{eq:decom0} completes the proof of \eqref{mom_m_r_llnr}.

Now, we consider the case $\alpha\in(0,1)$, which follows from the very same arguments. We apply \citet[Proposition 1]{Fav(13)}; see also \citet[Section 3.1]{Fav(13)} for the case $\alpha=0$. In particular, we have
\begin{align*}
\E[M_{r,n}]&=(1-\alpha)_{(r-1)}{n\choose r}\frac{(\lambda n+\alpha)_{(n-r)}}{(\lambda n+1)_{(n-1)}}=\frac{(1-\alpha)_{(r-1)}}{r!}\frac{\Gamma(n+1)}{\Gamma(n-r+1)}\frac{\Gamma(n(\lambda+1)+\alpha-r)\Gamma(\lambda n+1)}{\Gamma(\lambda n+\alpha)\Gamma(n(\lambda +1))}.
\end{align*}
From the asymptotic behaviour of the ratio of Gamma functions \citep[Equation 1]{Tri(51)}, as $n\rightarrow+\infty$
\begin{equation}\label{eq:krmean}
\E[M_{r,n}]=n\mathcal{M}_{\alpha,\lambda}(r)+O(1),
\end{equation}
where
\begin{displaymath}
\mathcal{M}_{\alpha,\lambda}(r)=\frac{(1-\alpha)_{(r-1)}}{r!}\lambda^{1-\alpha}(\lambda+1)^{\alpha-r},
\end{displaymath}
which completes the proof of the large $n$ behaviour of $\E[M_{r,n}]$ in \eqref{mom_m_r_llnr}. Still from \citet[Proposition 1]{Fav(13)},
\begin{align*}
\V[M_{r,n}]&=\left(\frac{\alpha(1-\alpha)_{(r-1)}}{r!}\right)^{2}(2r)!{n\choose 2r}\left(\frac{\lambda n}{\alpha}\right)_{(2)}\frac{(\lambda n+2\alpha)_{(n-2r)}}{(\lambda n)_{(n)}}\\
&\quad\quad+(1-\alpha)_{(r-1)}{n\choose r}\frac{(\lambda n+\alpha)_{(n-r)}}{(\lambda n+1)_{(n-1)}}\left(1-(1-\alpha)_{(r-1)}{n\choose r}\frac{(\lambda n+\alpha)_{(n-r)}}{(\lambda n+1)_{(n-1)}}\right)\\
&=\left(\frac{\alpha(1-\alpha)_{(r-1)}}{r!}\right)^{2}\frac{\Gamma(n+1)}{\Gamma(n-2r+1)}\left(\frac{\lambda n}{\alpha}\right)_{(2)}\frac{\Gamma(n(\lambda+1)+2\alpha-2r)\Gamma(\lambda n)}{\Gamma(\lambda n+2\alpha)\Gamma(n(\lambda+1))}\\
&\quad\quad+\frac{(1-\alpha)_{(r-1)}}{r!}\frac{\Gamma(n+1)}{\Gamma(n-r+1)}\frac{\Gamma(n(\lambda+1)+\alpha-r)\Gamma(\lambda n+1)}{\Gamma(\lambda n+\alpha)\Gamma(n(\lambda +1))}\\
&\quad\quad\quad\times\left(1-\frac{(1-\alpha)_{(r-1)}}{r!}\frac{\Gamma(n+1)}{\Gamma(n-r+1)}\frac{\Gamma(n(\lambda+1)+\alpha-r)\Gamma(\lambda n+1)}{\Gamma(\lambda n+\alpha)\Gamma(n(\lambda +1))}\right)\\
&=\left(\frac{\alpha(1-\alpha)_{(r-1)}}{r!}\right)^{2}\frac{\Gamma(n+1)}{\Gamma(n-2r+1)}\left(\frac{\lambda n}{\alpha}\right)_{(2)}\frac{\Gamma(n(\lambda+1)+2\alpha-2r)\Gamma(\lambda n)}{\Gamma(\lambda n+2\alpha)\Gamma(n(\lambda+1))}\\
&\quad\quad+\frac{(1-\alpha)_{(r-1)}}{r!}\frac{\Gamma(n+1)}{\Gamma(n-r+1)}\frac{\Gamma(n(\lambda+1)+\alpha-r)\Gamma(\lambda n+1)}{\Gamma(\lambda n+\alpha)\Gamma(n(\lambda +1))}\\&\quad\quad-\left(\frac{(1-\alpha)_{(r-1)}}{r!}\frac{\Gamma(n+1)}{\Gamma(n-r+1)}\frac{\Gamma(n(\lambda+1)+\alpha-r)\Gamma(\lambda n+1)}{\Gamma(\lambda n+\alpha)\Gamma(n(\lambda +1))}\right)^{2}.
\end{align*}
From the asymptotic behaviour of the ratio of Gamma functions \citep[Equation 1]{Tri(51)}, as $n\rightarrow+\infty$
\begin{align*}
\V[M_{r,n}]&=\left(\frac{\alpha(1-\alpha)_{(r-1)}}{r!}\right)^{2}\left(\frac{\lambda n}{\alpha}\right)_{(2)}A+\frac{(1-\alpha)_{(r-1)}}{r!}B-\left(\frac{(1-\alpha)_{(r-1)}}{r!}\right)^{2}C,
\end{align*}
where 
\begin{align*}
&A=(\lambda+1)^{2\alpha-2r}\lambda^{-2\alpha}\left(1+\frac{\lambda r(2+\lambda-2\lambda r)+\alpha-4\lambda r\alpha-2\alpha^{2}}{n(\lambda+1)\lambda}\right)+O(n^{-2})
\end{align*}
\begin{align*}
&B=n(\lambda+1)^{\alpha-r}\lambda^{1-\alpha}+O(n^{-1})
\end{align*}
and
\begin{align*}
&C=(\lambda+1)^{2\alpha-2r}\lambda^{2-2\alpha}n^{2}\left(1+2\frac{\lambda r(2+\lambda-\lambda r)+\alpha-2\lambda r\alpha-\alpha^{2}}{2n\lambda(\lambda+1)}\right)+O(n^{-2}),
\end{align*}
i.e.,
\begin{equation}\label{eq:krvar}
\V[M_{r,n}]=n\mathcal{S}^{2}_{\alpha,\lambda}(r)+O(1),
\end{equation}
where
\begin{align*}
\mathcal{S}^{2}_{\alpha,\lambda}(r)&=\frac{(1-\alpha)_{(r-1)}}{r!}(\lambda+1)^{\alpha-r}\lambda^{1-\alpha}+\left(\frac{(1-\alpha)_{(r-1)}}{r!}\right)^{2}(\lambda+1)^{2\alpha-2r}\lambda^{-2\alpha+2}\left(\frac{\alpha(\lambda+1)-(\lambda r+\alpha)^{2}}{\lambda(\lambda+1)}\right).
\end{align*}
The proof of the law of large numbers \eqref{mom_m_r_llnr} follows by an application of  \eqref{eq:krmean} and \eqref{eq:krvar}. In particular, we write
\begin{equation}\label{eq:decom}
\frac{M_{r,n}-n\mathcal{M}_{\alpha,\lambda}(r)}{n}=\frac{M_{r,n}-\E[M_{r,n}]}{n}+\frac{\E[M_{r,n}]-n\mathcal{M}_{\alpha,\lambda}(r)}{n}
\end{equation}
where, 
\begin{itemize}
\item[i)] from \eqref{eq:krmean}, as $n\rightarrow+\infty$
\begin{displaymath}
\frac{\E[M_{r,n}]-n\mathcal{M}_{\alpha,\lambda}(r)}{n}\rightarrow0;
\end{displaymath}
\item[ii)] from \eqref{eq:krvar}, for any $\varepsilon>0$ as $n\rightarrow+\infty$
\begin{displaymath}
\text{Pr}[|M_{r,n}-\E[M_{r,n}]|>n\varepsilon]\leq\frac{\V[M_{r,n}]}{n^{2}\varepsilon^{2}}=O(n^{-1}).
\end{displaymath}
\end{itemize}
Therefore, it holds $n^{-1}(M_{r,n}-\E[M_{r,n}])\stackrel{\text{p}}{\longrightarrow}0$ as $n\rightarrow+\infty$, which, according to \eqref{eq:decom} completes the proof of \eqref{mom_m_r_llnr}.

\section{Details on the variational inference algorithm}

\subsection{Full variational family}\label{app:vi_full_family}

Recall the definition of the ELBO
$$
\mathcal{L}(q) = E_q[\log p(\mathbf{x}, \bm z, \bm y^*, \bm v)] - E_q[\log q(\bm z, \bm y^*, \bm v)].
$$
where $q$ is as in \eqref{eq:vi_full}. Expanding terms explicitly, we have
\begin{align*}
\mathcal{L}(q) &= E_q[\log p(\mathbf{x} \mid \bm z, \bm y^*)] + E_q[\log p(\bm y^*)] + E_q[\log p(\bm z \mid \bm v)] + E_q[\log p(\bm v)] \\
& \quad - E_q[\log q(\bm v)] - E_q[\log q(\bm z)] - E_q[\log q(\bm y^*)].
\end{align*}

\begin{enumerate}
    \item Update for $q(y_{k\ell}^*)$: The optimal update for categorical distributions is
$$
\log \phi_{k\ell d} \propto \log \theta_{\ell d} + \sum_{i=1}^n r_{ik} \log \left[ (1-\beta_\ell) \mathbf{1}_{\{x_{i\ell}=d\}} + \beta_\ell\theta_{\ell x_{i\ell}} \right].
$$
Normalization is performed using log-sum-exp for numerical stability.

\item Update for $q(z_i)$:
The optimal cluster assignment update is:
$$
\log r_{ik} \propto E_q[\log \pi_k] + \sum_{\ell=1}^L \sum_{d=1}^{D_\ell} \phi_{k\ell d} \log \left[ (1-\beta_\ell) \mathbf{1}_{\{x_{i\ell}=d\}} + \beta_\ell \theta_{\ell x_{i\ell}} \right],
$$
where
$$
E_q[\log \pi_k] = \psi(a_k) - \psi(a_k + b_k) + \sum_{j=1}^{k-1}[\psi(b_j)-\psi(a_j+b_j)],
$$
with $\psi(\cdot)$ denoting the digamma function.

\item Update for $q(v_k)$:
The optimal Beta update parameters are given by:
$$
a_k = 1 - \alpha + \sum_{i=1}^n r_{ik}, \quad b_k = \theta + k\alpha + \sum_{i=1}^n\sum_{m=k+1}^K r_{im}.
$$
\end{enumerate}
These updates result directly from expectations involving stick-breaking construction and multinomial assignments, and standard algebraic manipulations of Beta distributions.

\subsection{Collapsed variational family}\label{app:vi_collapsed_family}

Recalling the derivation in the main paper, the objective function is the following lower bound on the evidence
\begin{align}
  \widetilde{\mathcal L}(q)
  &=
  \sum_{i=1}^{n}\sum_{\ell=1}^{L}
              \log\bigl(\beta_\ell\,\theta_{\ell,x_{i\ell}}\bigr)
  +
  \sum_{k=1}^{K}\sum_{\ell=1}^{L}
        \log f_{\mathrm{soft}}\bigl(\bm x_{k\ell}\bigr)
  \notag\\[4pt]
  &\quad
  +\sum_{i=1}^{n}\sum_{k=1}^{K}
       r_{ik}\Bigl[
         \psi(a_k)-\psi(a_k+b_k)
         +\sum_{j<k}
               \bigl\{\psi(b_j)-\psi(a_j+b_j)\bigr\}
       \Bigr]
  \notag\\
  &\quad
  +\sum_{k=1}^{K-1}\Bigl[
        (1-\alpha-1)\bigl\{\psi(a_k)-\psi(a_k+b_k)\bigr\}
        +(\theta+k\alpha-1)\bigl\{\psi(b_k)-\psi(a_k+b_k)\bigr\}
       \Bigr]
  \notag\\
  &\quad
  -\sum_{i=1}^{n}\sum_{k=1}^{K} r_{ik}\log r_{ik}
  -
  \sum_{k=1}^{K-1}
     \Bigl[
        \log B(a_k,b_k)
        -(a_k-1)\psi(a_k)
        -(b_k-1)\psi(b_k)
        +(a_k+b_k-2)\psi(a_k+b_k)
     \Bigr],
  \label{eq:JensenELBO_full}
\end{align}

We maximise $\widetilde{\mathcal L}(q)$ alternately in $q(\bm z)$ and $q(\bm v)$.

\begin{enumerate}
    \item Update for $q(\bm z)$ (responsibilities):
Keeping $q(\bm v)$ fixed, collect all terms in \eqref{eq:JensenELBO_full} that
depend on a single $z_i$:
\[
  \widetilde{\mathcal L}(q)
  = \text{const}
    +\sum_{k=1}^{K} r_{ik}\,
        \Bigl\{
          \E_{q(\bm v)}[\log\pi_k]
          +\sum_{\ell=1}^{L}
             \log\frac{f_{\text{soft}}(\bm x_{k\ell}^{+i})}
                          {f_{\text{soft}}(\bm x_{k\ell}^{-i})}
        \Bigr\}
        -\sum_{k} r_{ik}\log r_{ik},
\]
where $\,\bm x_{k\ell}^{\pm i}$ are the soft-count vectors with record
$i$ included/excluded.  Enforcing $\sum_k r_{ik}=1$ by a Lagrange multiplier
and exponentiating gives
\begin{equation}
  \log r_{ik}
  =
     \psi(a_k)-\psi(a_k+b_k)
                 +\sum_{j<k}[\psi(b_j)-\psi(a_j+b_j)]
     +\sum_{\ell=1}^{L}
        \log\frac{f_{\text{soft}}(\bm x_{k\ell}^{+i})}
                  {f_{\text{soft}}(\bm x_{k\ell}^{-i})}
     -\log Z_i,
  \label{eq:r_update}
\end{equation}
where $Z_i$ is the normalizing constant.  

\item Update for $q(\bm v)$: Holding $q(\bm z)$ fixed, differentiate $\widetilde{\mathcal{L}}(q)$ in
$a_k$ and $b_k$; since the terms in which they appear mirror those of
\citet{Blei06} we quote the closed forms:
\begin{equation}
  a_k^{\text{new}} = 1-\alpha + N_k,
  \qquad
  b_k^{\text{new}} = \theta + \alpha(k-1) + \sum_{j>k}N_j,
  \qquad
  N_k=\sum_{i=1}^{n}r_{ik}.
  \label{eq:ab_update}
\end{equation}
The digamma expectations used in \eqref{eq:r_update} are then updated
accordingly.

\end{enumerate}

\subsection{Updates for $\lambda$ and $\alpha$}\label{app:hyperparams}

Gradients of $\widetilde{\mathcal L}(q)$ with respect to $\lambda$ and $\alpha$ are
\[
  \frac{\partial\widetilde{\mathcal L}}{\partial\lambda}
     = n\sum_{k} \Bigl[\psi(b_k)-\psi(a_k+b_k)\Bigr],
  \qquad
  \frac{\partial\widetilde{\mathcal L}}{\partial\alpha}
     = \sum_{k} \Bigl[\psi(a_k)-\psi(a_k+b_k)
                      -k\bigl(\psi(b_k)-\psi(a_k+b_k)\bigr)\Bigr],
\]
with corresponding Hessian elements derived via trigamma;
one damped Newton step per outer iteration suffices and preserves ELBO
monotonicity.

\subsection{Stochastic Variational Inference}\label{app:svi}

We now show how the full–batch variational-inference scheme \ref{app:vi_collapsed_family} translates into an efficient stochastic variational–inference (SVI) algorithm \citep{hoffman2013stochastic}.  
Throughout,  $B\ll n$  denotes the mini-batch size and $\mathcal B\subseteq\{1,\dots,n\}$ the index set of the current mini-batch.

Let $\mathbf n=(n_{k\ell d})$ collect the global soft counts and $\boldsymbol\gamma=(a_k,b_k)$ the stick–breaking natural parameters.
For a single mini-batch we form the stochastic ELBO estimator
\begin{equation}
  \widehat{\mathcal L}(\mathcal B)
  =
  \frac{n}{B}
  \sum_{i\in\mathcal B}
      \sum_{k=1}^{K}
         r_{ik}\,\xi_{ik}
  +  \mathbb E_{q(\bm v)}[\log p(\bm v)]
     -\mathbb E_{q(\bm v)}[\log q(\bm v)],
  \label{eq:stochastic_elbo}
\end{equation}
where $\xi_{ik}$ is the same per-record contribution used in the
full-batch ELBO~\eqref{eq:JensenELBO_full}.  The factor $n/B$ makes
$\widehat{\mathcal L}$ an unbiased estimate of the full ELBO.
Define scaled counts
\begin{equation}
  \hat n_{k\ell d} = \frac{n}{B}\,
  \sum_{i\in\mathcal B}
     r_{ik} \mathbf 1\{x_{n\ell}=d\},
  \qquad
  \hat N_k =\sum_{\ell,d}\hat n_{k\ell d},
  \label{eq:scaled_counts}
\end{equation}
and the corresponding stick parameters
$\hat a_k=1-\alpha+\hat N_k$, $\hat b_k=\theta+\alpha(k-1)+\sum_{j>k}\hat N_j$ obtained exactly as in
\eqref{eq:ab_update}.  
By construction $\mathbb E[\hat n_{k\ell d}]=n_{k\ell d}$ and $\mathbb E[\hat a_k]=a_k^\star$, so the updates below are unbiased.

Writing $\mathbf n^{(t)}$ and $\boldsymbol\gamma^{(t)}$ for the global
variational parameters at iteration $t$, a natural-gradient ascent step
\citep{hoffman2013stochastic} with step size $\rho_t=(t_0+t)^{-\kappa}$, $\,\kappa\in(0.5,1]$, yields
\begin{align}
  \mathbf n^{(t+1)}      &= (1-\rho_t)\,\mathbf n^{(t)}
                           +\rho_t\,\hat{\mathbf n}(\mathcal B),
                           \label{eq:svi_n_update}\\
  \boldsymbol\gamma^{(t+1)} &= (1-\rho_t)\,\boldsymbol\gamma^{(t)}
                           +\rho_t\,\hat{\boldsymbol\gamma}(\mathcal B),
                           \label{eq:svi_gamma_update}
\end{align}
where $\hat{\mathbf n}$ and $\hat{\boldsymbol\gamma}$ are obtained from
\eqref{eq:scaled_counts}.  This recursion is a Robbins–Monro
stochastic-approximation scheme.

In our experiments, we use $t_0 = 1$, $\kappa = 0.9$.


\end{document}